\documentclass[12pt]{article}

\usepackage[left=15mm,top=15mm,right=15mm,bottom=24mm,nohead,nofoot]{geometry}

\usepackage{times}

\usepackage{epsfig}

\title{Change in Recessive Lethal Alleles Frequency in Inbred Populations}

\author{Arindam RoyChoudhury \vspace{-0.1in}\\
\footnotesize
Department of Biostatistics, Columbia University, New York NY 10032, U.S.A.,\vspace{-0.2in}\\ 
\footnotesize
Email: ar2946@columbia.edu
\footnotesize
Phone: 1-212-342-1268, FAX: 1-212-305-9408
}

\date{}

\renewcommand{\baselinestretch}{2}

\begin{document}

\maketitle

\begin{abstract}
In a population practicing consanguineous marriage, rare recessive lethal alleles (RRLA) have higher chances of affecting phenotypes. As inbreeding causes more homozygosity and subsequently more deaths, the loss of individuals with RRLA decreases the frequency of these alleles. Although this phenomenon is well studied in general, here some hitherto unstudied cases are presented. An analytical formula for the RRLA frequency is presented for infinite monoecious population practicing several different types of inbreeding. In finite diecious populations, it is found that more severe inbreeding leads to quicker RRLA losses, making the upcoming generations healthier. A population of size 10,000 practicing 30\% half-sib marriages loses more than 95\% of its RRLA in 100 generations; a population practicing 30\% cousin marriages loses about 75\% of its RRLA. Our findings also suggest that given enough resources to grow, a small inbred population will be able to rebound while losing the RRLA.
\end{abstract}

\section{Introduction}

Consanguineous marriage (CM), the marriage between two related individuals, is commonly practiced in many populations. At least 20\% of human populations live in cultures with preference for such marriages \cite{model.darr2002}. A birth study in Birmingham, UK found that 57\% of all couples in British Pakistani population were first cousins \cite{model.darr2002}. Half-sib marriages were allowed in ancient Greece, whereas full sib marriages were fairly common in Roman Egypt \cite{hopkins1980}. About one in six marriages in first-third century CE  Roman Egypt are recorded to be between full sibs \cite{bagnall.frier2006}. On the other hand, non-deliberate inbreeding may occur in a population with small size. Inbreeding resulting from CM is known to increase the chances of still-births and infant deaths \cite{bittles.neal.1994,stoltenberg.et.al.1999,stoltenberg.et.al.1999b}. 

Inbreeding increases number of homozygous individuals (with two copies of a recessive lethal allele), and as a result the homozygous individuals die due to selection. As a result, the frequency of such alleles reduces among the child generation (see, for example, \cite{werren1993,henter2003,charlesworth.willis2009}). It is known that such alleles are present in many (human or wild) populations \cite{charlesworth.willis2009}.
Here we study the amount of loss of rare recessive lethal alleles (RRLA) under different types of inbreeding, in an infinite monoecious population and in finite diecious populations with various parameters.

\section{Methods}

\subsection{Infinite monoecious population}

Let us consider an infinite monoecious population with discrete generations. Also assume that at a certain locus there are two possible alleles ``A" and ``a" and the effect of mutation is negligible. Suppose that ``a" is a rare recessive lethal allele. Let $p_t$ $(1> p_t >0)$ be the ``a''-frequency (frequency of ``a'') at Generation $t$ after any deaths have occurred due to selection. Also assume that the population practices $k$ different types of inbreeding $I_1, I_2, \dots, I_k$ with inbreeding coefficients $\gamma_1, \gamma_2, \dots, \gamma_k$ respectively. A fraction $\alpha_i$ $(>0)$ of the mating are of type $I_i$. (The remaining fraction of $1-\sum_{i=1}^k \alpha_i \, (>0)$ matings are random.) In inbreeding type $I_i$, the most recent common ancestor is $t_i$ generations before the generation of the child. Thus, the probability that a child (at Generation 0) of $I_i$ will be ``aa''  is 
\begin{eqnarray}
\gamma_i \, p_{-t_i} + (1-\gamma_i) \, p_{-1}^2.
\end{eqnarray}
If all the children in Generation 0 had survived, then the frequency of ``aa" individuals would have been:
\begin{eqnarray}
b_0 = \mbox{Pr(``aa'') } & = & \mbox{Pr(``aa''  } \vert \mbox{ parents' mating was random } ) \, \mbox{Pr}(\mbox{ parents' mating was random } ) \nonumber \\
&     &  \, + \sum_{i=1}^k \mbox{Pr(``aa'' } \vert \mbox{ parents' mating was of type } I_i) \, \mbox{Pr}(\mbox{ parents' mating was of type } I_i) \nonumber \\
& = &  \mbox{Pr(``aa''} \vert \mbox{ parents' mating was random } ) \, \left(1-\sum_{i=1}^k \alpha_i \right) \nonumber \\
&     &  \, + \sum_{i=1}^k \mbox{Pr(``aa'' } \vert \mbox{ parents' mating was } I_i) \, \alpha_i \nonumber \\
& = &  p_{-1}^2 \, \left(1-\sum_{i=1}^k \alpha_i \right) +  \sum_{i=1}^k  \, \left(\gamma_i \, p_{-t_i} + (1-\gamma_i) \, p_{-1}^2\right) \, \alpha_i \nonumber \\
& = & p_{-1}^2 \, (1-\sum_{i=1}^n \alpha_i \, \gamma_i) +  \sum_{i=1}^n \alpha_i \, \gamma_i \,  p_{-t_i} \label{eq:b0}
\end{eqnarray}
Noting that $\alpha_i > 0$, $1 > \gamma_i > 0$ and  $1-\sum_{i=1}^k \alpha_i >0$, it follows that $1-\sum_{i=1}^n \alpha_i \, \gamma_i > 0$ and subsequently both sides of Eq. (\ref{eq:b0}) is positive.
The ``a''-frequency before any death (i.e. in the gamete population) is $p_{-1}$. This is because there is no change in allele frequency between the parent and gamete populations.
As ``aa'' individuals die, each taking away two ``a'' alleles from the population, ``a''-frequency at Generation 0 becomes
\begin{eqnarray}
p_0 = \frac{p_{-1} - 2 \, b_0}{1 - 2 \, b_0} \label{eq:p0}
\end{eqnarray}
As $b_0 > 0$, it follows that $p_{-1} > p_0$, and thus the allele-frequency has decreased. Note that even without any deliberate inbreeding there will be some homozygosity, resulting in death due to selection.

If inbreeding continues, the frequency of lethal alleles keeps decreasing. To see the long-term effect of this in \textit{finite diecious} populations with mutation, we use computer simulations. 

\subsection{Simulation: finite diecious population}

We simulated from a Wright-Fisher model. The populations have discrete generations with fixed generation size: $N$ (even number) individuals with $N/2$ couples ($N/2$ females and $N/2$ males). Each individual has two alleles at $L$ independently segregating loci. The rare recessive lethal alleles (RRLA) are present among these $L$ loci with a predetermined initial frequency. (The allele frequency is allowed to vary across the $L$ loci, but the fraction among the total $L$ loci is fixed.)
At each new generation, each child is produced by first randomly selecting a parental couple from the previous generation, and then taking one (randomly selected) allele from each locus of each parent. 
We also force a pre-determined fraction of parental couples to be randomly selected from a set of related couples (sibs, half-sibs or cousins, according to predetermined parameters) so that their children 
are inbred. However, if at least one parent has two copies of the lethal allele 
(at any locus) then we do not generate a child from them and instead select another couple with same degree of relatedness. Once the children are generated, we mutate a very small fraction (1 in $10^{8}$) of 
their healthy alleles to lethal alleles to simulate the effect of mutation. This is done by changing a healthy allele of a random individual to lethal. 

\section{Results}

\begin{figure}[h]
\centering
\epsfig{file={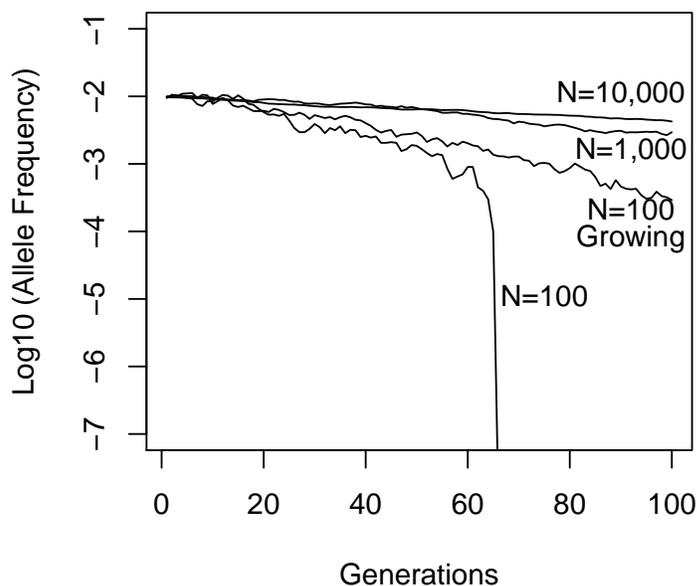},width={4 in},height={4 in}}
\caption{Allele frequencies over generations in populations with no deliberate inbreeding and initial allele frequency $= 1$\%.}
\label{fig:inb0}
\end{figure}

We first compare the recessive lethal allele-frequencies of four simulated populations with no 
deliberate inbreeding (Figure \ref{fig:inb0} ). One of the populations has a 2\% growth rate per generation. Rest have constant sizes.
The RRLA frequencies in populations with N=10,000 and 1,000 go down by 50\% and 70\% respectively. There is more rapid decline in RRLA frequency in the populations with N=100. The decline is more in the population with constant size as it loses all its RRLA within 70 generations. However, even the growing population lost more than 97\% percent of its RRLA in 100 generations.

\begin{figure}[h]
\centering
\epsfig{file={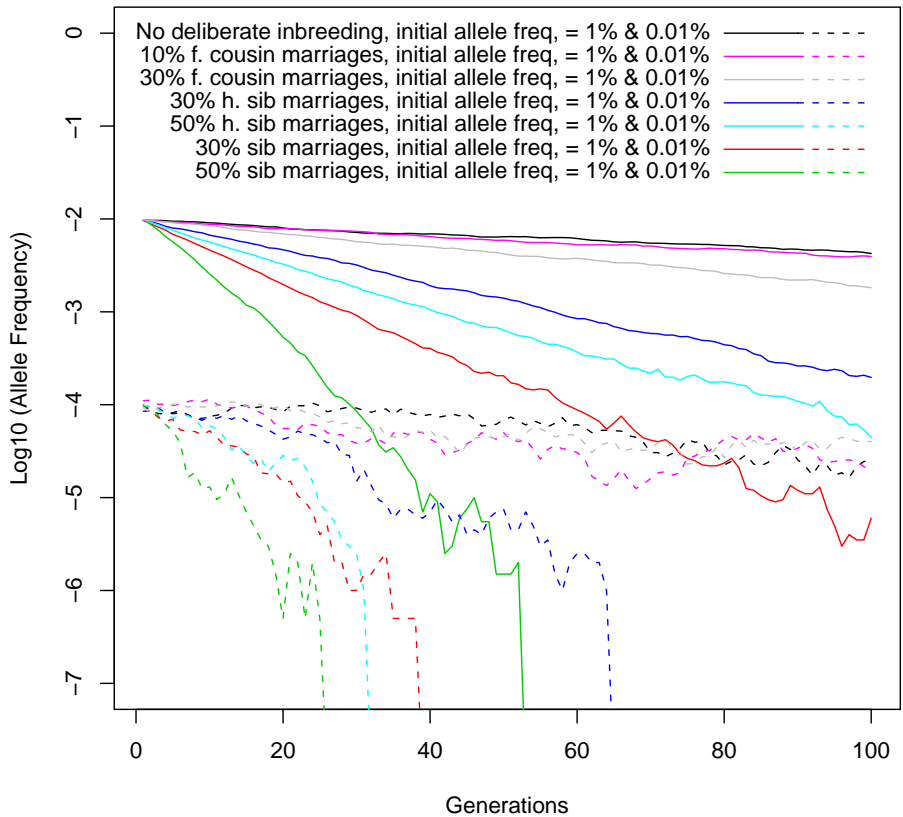},width={5 in},height={5 in}}
\caption{Allele frequencies over generations in populations with constant size of 10,000.}
\label{fig:N10000}
\end{figure}

Then we compare the frequencies in populations with constant size of 10,000 and different type of 
inbreeding and different initial frequency of RRLA (Figure \ref{fig:N10000}). For the initial frequency of 0.01\%, 
the populations with sib and half-sib matings lost the RRLA completely within 100 generations. 
Also, in the population with 1\% initial frequency and 50\% half-sib matings, the alleles were lost within 
100 generations. The rest, although did not lose RRLA completely, showed significant decline. 
The population with 30\% half-sib marriages loses more than 95\% of its RRLA; the population with 30\% cousin marriages loses about 75\% of its RRLA; however, the population with 10\% half-sib marriages retains about two-third of its RRLA. 

\begin{figure}[h]
\centering
\epsfig{file={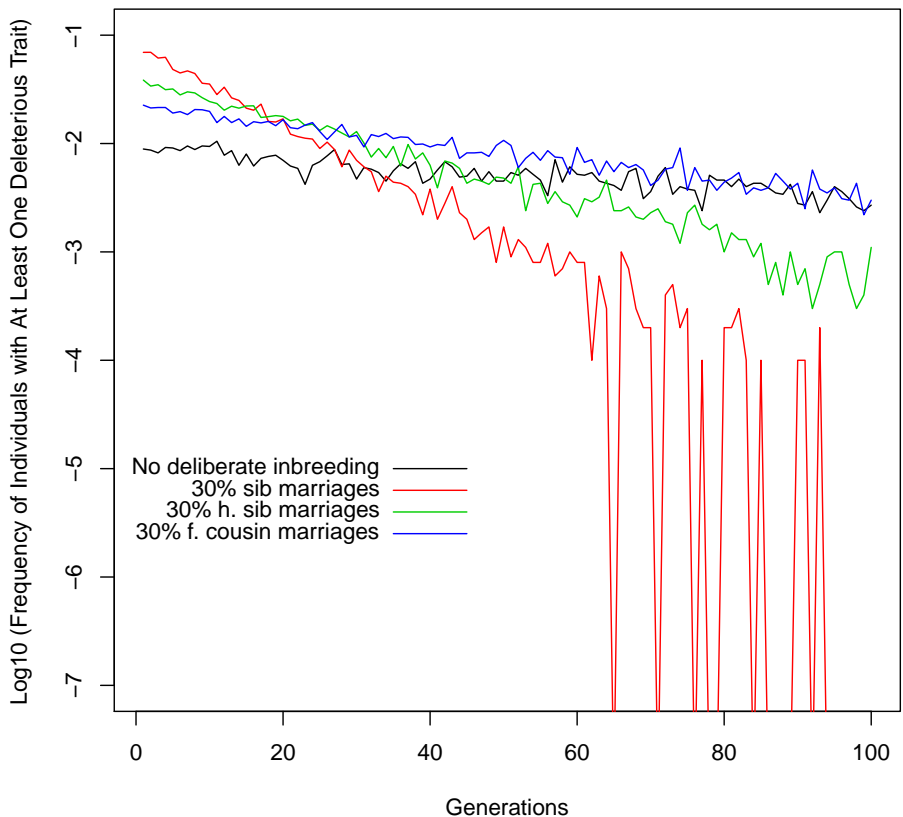},width={5 in},height={5 in}}
\caption{Proportion of individual with lethal traits in  populations with constant size of 10,000 and initial allele frequency $= 1$\%.}
\label{fig:ffp}
\end{figure}

Next we compare the proportion of individual with lethal traits across populations with 
constant size of 10,000, initial allele frequency of 1\% and different type of inbreeding (Figure \ref{fig:ffp}). The
proportion of individuals with lethal traits (that is, ``aa'' genotype before any deaths) declines most rapidly for sib-matings, and little less rapidly for half-sib matings. The population with 30\% full-sib marriages stops having the lethal traits within 95 generations. This is because the RRLA frequency in this population is so low at that point, that it is unlikely to have a homozygous individual.

\renewcommand{\baselinestretch}{1}

\renewcommand{\baselinestretch}{2}

\section{Discussion}

Inbreeding alters the genotype frequencies, which results in removal of lethal alleles in homozygous individuals due to selection.
Based on this basic fact, we presented three major findings. First,  even a very small population has chances of surviving the effects 
of inbreeding and rebounding. As Figure \ref{fig:inb0} shows, both the populations that started with size $N=100$ 
(one constant size and the other growing) lost the RRLA rapidly.  In the first few generations there were many individuals with lethal traits, 
which facilitated the loss of the RRLA, and made the upcoming generations healthier. The high level of RRLA did not stop the population 
from growing in size either. Thus, small endangered wild populations that are at high risk of inbreeding has the possibility of rebounding.

The second major result focuses on populations that actively practices inbreeding. As shown in Figures \ref{fig:N10000} and \ref{fig:ffp},   more severe the inbreeding, quicker the loss of RRLA. That is, the populations with higher rate of inbreeding and with more acute type of inbreeding will lose the RRLA more rapidly. We have also produced the rate of RRLA loss for different scenarios. One of our simulations shows that a population of size 10,000 practicing 30\% cousin marriages loses about 75\% of its RRLA in 100 generations. This is significant as there are many populations practicing 30\% or more cousin marriages. The Figure \ref{fig:ffp} shows that the proportion of individuals with lethal traits declines (or disappears) after a steady period of inbreeding. A similar result is also shown with populations practicing half-sib marriages.

In addition to the above we have also provided a formula for RRLA-frequency in an infinite monoecious population practicing several different types of inbreeding. The formula is given as a function of the RRLA-frequencies of the previous generations. The formula also depends on the inbreeding coefficients of the types of inbreeding practiced, as well as on their frequencies (which are assumed to be constant across generations). Note that this formula may be used to approximate the RRLA frequency in large diecious populations; this formula is especially useful in population where several different types of inbreeding are practiced.

\bibliographystyle{plain}
\bibliography{inbreeding.3}

\end{document}